\documentclass[aps,superscriptaddress,fleqn,nofootinbib]{revtex4}
\usepackage{amsmath,amssymb}
\usepackage{graphicx,color}

\def\bm#1{\mbox{\boldmath $#1$}}

\begin{document}

\title{Synchronization Engineering: Theoretical Framework and
Application to Dynamical Clustering} \author{Hiroshi Kori}
\thanks{present address: Division of Advanced Sciences,
Ochadai Academic Production, Ochanomizu University, Tokyo, 112-8610,
Japan. Electronic address: kori.hiroshi@ocha.ac.jp}
\affiliation{Department of Mathematics, Hokkaido University, Sapporo,
Hokkaido, 060-0810, Japan} \author{Craig G. Rusin}
\affiliation{Department of Chemical Engineering, University of Virginia,
Charlottesville, Virginia 22904-4741, USA} \author{Istv\'{a}n Z. Kiss}
\affiliation{Department of Chemistry, Saint Louis University, St. Louis,
Missouri 63103, USA} \author{John L. Hudson} \affiliation{Department of
Chemical Engineering, University of Virginia, Charlottesville, Virginia
22904-4741, USA}

\begin{abstract}
A method for engineering the behavior of populations of rhythmic
elements is presented.  The framework, which is based on phase models,
allows a nonlinear time-delayed global feedback signal to be constructed which
produces an interaction function corresponding to the desired 
behavior of the system.  It is shown theoretically and confirmed in
numerical simulations that a polynomial, delayed feedback is a versatile
tool to tune synchronization patterns.  Dynamical states consisting of
one to four clusters were engineered to demonstrate the application of
synchronization engineering in an experimental electrochemical system.
\end{abstract}

\maketitle

{\bf Populations of interacting rhythmic components can produce complex
behavior in biology \cite{winfree80,tass99}, communications
\cite{carroll96}, population dynamics \cite{blasius99}, and chemistry
\cite{kuramoto84,ertl91,epstein98,mikhailov06}. In biology,
synchronization can be beneficial, such as in orchestrating the
circadian rhythms in mammals, or pathological, such as in the occurrence
of Parkinson's disease.  We consider here the engineering of desirable
states through the introduction of mild feedback, mild such that the
behavior of the individual components is not substantially changed by
the introduction of the external signal. In a previous paper
\cite{kiss07}, we have experimentally demonstrated a general methodology
for engineering a target dynamical behavior in oscillator assemblies.
The aim of the present paper is to describe the theory behind our
methodology and to verify it by numerical and experimental studies.}

\section{Introduction}
Ensembles of self-sustained oscillators can spontaneously organize their
collective dynamical behavior as a result of interaction among
elements.  Examples can be found in biological \cite{winfree80,tass99},
chemical \cite{kuramoto84,ertl91,epstein98,mikhailov06}, and ecological
systems \cite{blasius99}, communications \cite{carroll96}, as well as
human activities \cite{neda00}.  Global behaviors such as
synchronization are often responsible for the formation of certain
beneficial biological functions, such as orchestrating the sleep/wake
cycle (circadian rhythm) of mammals \cite{reppert02}. Conversely,
pathological synchronization may induce serious problems, e.g., tremors
in Parkinson's disease \cite{tass99} and abnormal sway in London's
Millennium Bridge \cite{strogatz05}.

Interactions involving feedback among rhythmic elements are often
associated with the formation of dynamical order. In many cases,
feedback plays an essential role in sustaining dynamical stability and
in suppressing complexity.  In chemical systems, several types of
complex synchronized behavior emerge by introducing global feedback,
which otherwise shows simple synchronization or chemical turbulence
\cite{vanag00,kim01,mikhailov06}.  It may be expected that feedback
loops among neuronal clusters contribute to the design of
the complex dynamical functionality of the brain \cite{manrubia04}.  In
medical applications, heart pacemakers, deep brain pacemakers, and
feedback control techniques have been proposed to eliminate pathological
synchronization \cite{christini02,tass99}.  For applications which
involve biological neurons, a mild control is desired to avoid side
effects and to maintain the fundamental nature of neurons in the system.

In this paper we present a comprehensive theory for designing the
collective dynamics of a rhythmic population using external feedback
and, as an application of its use, demonstrate the power of the
methodology for creating various cluster states in both numerical
simulations and experimental studies.

This method has been shown to be extremely robust in engineering
collective dynamical behavior in electrochemical experiments
\cite{kiss07}.  Our method utilizes phase modeling \cite{kuramoto84} (as
opposed to a physiochemical modeling) to describe the dynamical behavior
of rhythmic elements.  The simplicity and analytical tractability of the
phase model is exploited to design an optimal, delayed,
nonlinear feedback signal for obtaining a desired collective behavior.
The only properties required to construct the phase model are the
waveform and the phase response function of the oscillator, which can be
experimentally measured. Sec. II outlines the mathematics of the
methodology. In Sec. III, the detailed description of the feedback
design method is presented, which is further developed in Sec. IV for
both harmonic and slightly inharmonic waveforms.  Secs. V and VI present
examples of synchronization engineering using numerical studies via the
Brusselator model and experimental studies using electrochemical
oscillators, respectively.

\section{general methodology}
Our methodology seeks to engineer a target collective
dynamical behavior in a population of limit-cycle oscillators through
feedback.  A phase model is used to describe the collective behavior of
a population of weakly coupled oscillators.  For $N$ oscillators with
general interactions (which also admits interactions through a complex
network), the phase model is given as
\begin{equation}
 \frac{d \phi_i}{dt} = \omega_i + K \sum_j H_{ij} (\phi_j - \phi_i),
  \label{pm_general}
\end{equation}
where $\phi_i$ ($0\le \phi_i < 2\pi$) and $\omega_i$ are the phase and
the natural frequency of the oscillator $i$, $K$ is a coupling strength
($K>0$), and the $2\pi$ periodic function $H_{ij}(\phi)$ is the (phase)
interaction function (or, phase coupling function). The phase model is
derived as the first order approximation of a coupled limit-cycle
oscillator system, where the small quantity is the coupling intensity
$K$ \cite{kuramoto84}. The interaction function $H_{ij}(\phi)$ can be
calculated from the properties of the limit-cycle oscillator $i$ and
physical interaction from the oscillator $j$ to the oscillator $i$. The
details of this derivation are presented in Sec.
\ref{sec:theory}. Although, as presented in Sec. \ref{sec:theory}, our
proposed theory may deal with Eq. (\ref{pm_general}), we mostly devote
ourselves in the present paper to the case of global feedback.
In such a case, the phase model is reduced to
\begin{equation}
 \frac{d \phi_i}{dt} = \omega_i + \frac{K}{N} \sum_j H_i (\phi_j - \phi_i).
  \label{pm-hetero}
\end{equation}
Moreover, if the heterogeneity is sufficiently small compared to the
feedback intensity $K$, we can treat the system as identical
oscillators. In such a case, we may use the following phase model:
\begin{equation}
 \frac{d \phi_i}{dt} = \omega + \frac{K}{N} \sum_j H (\phi_j - \phi_i).
  \label{pm}
\end{equation}
For simplicity, we outline our methodology in terms of Eq.~(\ref{pm}) in this
section.

In the phase model (\ref{pm}), dynamical evolution of the system
is predicted if the interaction function $H(\phi)$ and an initial
condition are given. This comes from the fact that we may set
$\omega=0$ and $K=1$ without loss of generality (using a rotating
reference frame, $\phi-\omega t \to \phi$ , and rescaling time, $Kt \to
t$). The relationship between the shape of the interaction function and
the global dynamical behavior of the system has been well studied. For
example, the conditions which admit perfect synchrony, perfect
desynchrony (the spray state) \cite{kuramoto84}, phase clustering
\cite{okuda93}, and slow switching dynamics
\cite{hansel93,kori01,kori03} are known. While it is possible for an
interaction function to admit multiple attractors, it is preferable to
have a single stable attractor (or at least a single dominant basin of
attraction).  If a coupled limit-cycle oscillator system has a phase
interaction function which results in a single stable attractor, the
system should exhibit the expected global behavior under general initial
conditions.

Two steps are required to engineer a desired target behavior:
\begin{itemize}
 \item[(i)] find an interaction function $H(\phi)$ which uniquely
	    stabilizes the desired collective behavior
 \item[(ii)] seek the physical feedback parameters which result in the
	    interaction function found in (i)
\end{itemize}
The difficulty of the step (i) depends on the desired target behavior.
To illustrate the engineering methodology, we have selected a simple
collective behavior which has been well characterized in terms of the
phase model: perfect synchrony and balanced cluster states.  Two
different approaches can be utilized to optimize feedback parameters.
The first approach simply requires knowledge of the {\em precise}
interaction function to be targeted.  However, there are many cases
(such as phase clustering) in which there exists a large family of valid
functions, each capable of producing the desired target behavior.  By
arbitrarily selecting one of these functions, the most effective means
of generating the target behavior may be overlooked.  Additionally, an
arbitrary choice of an interaction function can substantially increase
the difficulty of the feedback parameter optimization.  Therefore,
instead of targeting a precise interaction function, we place
constraints on specific properties of the interaction function as
required to generate the appropriate behavior.  The optimum feedback
parameters are selected such that the associated interaction function
meets these constraints with the minimum feedback amplitude.

To engineer a target interaction function into a physical system, we
introduce a feedback $K p(t)$ to some global parameter of the system
with the following functional form:
\begin{equation}
 p(t) =  \frac{1}{N} \sum_{i=1}^N h(x_i),
  \label{p}
\end{equation}
\begin{equation}
 h(x_i) = \sum_{n=0}^S k_n \{x_i(t-\tau_n)-a_0\}^n,
  \label{h}
\end{equation}
where $x_i(t)$ is an observable variable of the oscillator $i$ at time
$t$, $a_0$ is the time average of $x_i$, $k_n$ and $\tau_n$ the gain and
the delay of the $n^{\rm th}$ order feedback respectively, and $K$ and
$S$ the overall gain and the overall order of the feedback respectively.
Our choice of function (\ref{h}) is motivated from the fact
that each feedback term yields different combinations of intensities of
Fourier components and the feedback delay value $\tau_n$ controls the
ratio between the symmetric and antisymmetric Fourier components.  In
addition, the $n^{\rm th}$ harmonic of the interaction function is
efficiently enhanced by the $n^{\rm th}$ order feedback. Thus, flexible
and efficient design of the interaction function is possible.  In
Sec. \ref{sec:theory}, we show that any target interaction function
which is composed of $S^{\rm th}$ and lower Fourier components can be
indeed produced by using the $S$ overall order feedback. In particular,
when the waveform $x_i(t)$ is exactly harmonic, feedback parameter
values $\{k_n\}$ and $\{\tau_n\}$ may be calculated analytically, as
illustrated in Sec. \ref{sec:harmonic}.  For a general waveform, a
numerical optimization is often required to determine the feedback
parameters.

\section{theory of feedback design} \label{sec:theory}
We present a theory for designing external feedback signal
yielding a desired phase interaction function. Because the extension to
more complex situations is straightforward, it is suitable to start with
the case where the oscillator 1 is affected by the feedback signal
composed as a function of the state variable of the oscillator 2. The
dynamical equation for the oscillator 1 is then given as
\begin{equation}
  \frac{d \bm{A}_1 (t)}{dt} = \bm{F}_1 [\bm{A}_1(t)]  + K \bm{P}(t),
   \label{full_equation}
\end{equation}
where $\bm{A}_i =(x_i,y_i,\ldots)^{\rm T}$ is the state variable of the
oscillator $i$ $(i=1,2)$, $\bm{F}_i(\bm {A}_i)$ is a nonlinear function
admitting limit-cycle oscillation, and $\bm{P}(t)$ is a feedback
signal. The dynamical equation for the oscillator 2 is arbitrary
provided that it produces nearly periodic dynamics. We define the
observable variable to be $x$ and the variable to be perturbed by the
feedback to be $y$ (these variables need not be mutually exclusive,
which is the case in the numerical studies and experiments). Thus,
\begin{equation}
 \bm{P}(t) = \{0,h(x_2), 0, \ldots, 0\}^{\rm T},\label{P}
\end{equation}
where $h(x)$ is Eq. (\ref{h}).

Because we are interested in mild engineering, we assume that the
overall gain $K$ is small such that the dynamical behavior of the system
(\ref{full_equation}) can be approximated by the phase oscillator model,
\begin{equation}
 \frac{d \phi_1}{dt} = \omega_1 + K \sum_j H (\phi_2 - \phi_1).
  \label{pm_2}
\end{equation}
The interaction function $H(\phi_2-\phi_1)$ is computed as
\begin{equation}
 H(\phi_2-\phi_1) = \frac{1}{2\pi}\int_{0}^{2\pi} Z(\phi_1+\theta)
  h(\phi_2+\theta) d\theta
  \label{H}
\end{equation}
where $Z(\phi_1)$ and $h(\phi_2)$ are evaluated from single isolated
oscillators $1$ and $2$, respectively.
The function $Z(\phi_1)$ is referred to as the phase response function
(or, the phase sensitivity function) of the oscillator 1, which is the
gradient of the phase along the $y$-direction on the limit-cycle orbit
$\bm{A}_1^C(\phi)$,
\begin{equation}
 Z(\phi_1) = \left. \frac{\partial \phi_1}{\partial y_1} \right|_{\bm{A_1}=\bm{A}_1^C}.
\end{equation}  
There are several ways to measure $Z(\phi)$ of a given oscillator
\footnote{For example, see Ref.~\onlinecite{kuramoto84} for the
analytical derivation, the software by Ermentrout, xppaut, for the
numerical derivation, and Refs.~\onlinecite{galan05,kiss05,tsubo07} for
the experimental derivation. Some of the methods are reviewed in
Ref.~\onlinecite{izhikevich07}.}.  The function $h(\phi_2)$ is obtained
by first describing $x_2(t)$ as the function of the phase of the
oscillator, $x_2(\phi_2)$.  Because $\phi_2(t-\tau)=\phi_2(t)-\omega_2
\tau$ when the interaction is absent, we have
$x_2(\phi_2(t-\tau))=x_2(\phi_2(t)-\omega_2 \tau)$. As a result,
$h(\phi_2)$ assumes the form
\begin{equation}
 h(\phi_2) = \sum_{n=0}^S k_n \{x(\phi_2-\omega_2 \tau_n)-a_0\}^n,
\end{equation}
or, upon expanding $x(\phi)=\sum_l a_l e^{-il\phi}$, as
\begin{equation}
 h(\phi_2) = \sum_{n=0}^{S} k_n \left\{ \sum_{l\neq0} a_l
			e^{-il\phi_2}e^{il\omega_2 \tau_n} \right\}^n.
 \label{h_expansion}
\end{equation}
Thus, the phase model is an autonomous system despite the existence of
time delays in the original system (\ref{full_equation}).  Such an
approximation is valid so long as $K \tau$ remains small (note that the
dimension of $K$ is inverse time) \cite{izhikevich98,kori01}.

For given $Z(\phi_2)$, the feedback parameters ${k_n}$ and
${\tau_n}$ yielding a target $H(\phi)$ are found in the following way.
To simplify the problem, the functions are expanded into their
Fourier series.  $H(\phi) = \sum_l H_l e^{-il\phi}$, $Z(\phi) = \sum_l
Z_l e^{-il\phi}$ and $h(\phi) = \sum_{l} h_l e^{-il\phi}$ (where
$H_l=H_{-l}^*, Z_l=Z_{-l}^*$, and $h_l=h_{-l}^*$).
Using these Fourier coefficients, we obtain the relation
\begin{equation}
 H_l = h_l Z_{-l},
  \label{convolution}
\end{equation}
where $h_l$ is the function of ${k_n}$ and ${\tau_n}$. By solving a set
of complex equations (\ref{convolution}), the feedback
parameters ${k_n}$ and ${\tau_n}$ can be determined. In theory, any
interaction function composed of harmonic components $H_l$ for $0 \leq l
\leq S$ can be constructed using a feedback signal with an overall order
of $S$, provided that $z_l$ for $l=0,\ldots, S$ is
non-vanishing. 


It is important to point out that the flexibility of our engineering
method is reduced for certain types of oscillators. For example, the
Stuart-Landau oscillator (i.e., the normal form for the Hopf
bifurcation) has $Z_l=0$ for $l \ge 2$ \cite{kuramoto84}, forcing all
higher harmonics ($l \ge 2$) in the interaction function to vanish
regardless of the nonlinear terms in the feedback.  A similar problem
may occur in systems which contain special symmetry.  For example, the
Van der Pol oscillator has symmetry with respect to the center of the
oscillation.  This fact implies that $Z(\phi+\pi)=Z(\phi)$, i.e.,
$Z_l=0$ for even $l$.  Thus, the even Fourier components of the
interaction function vanish.  In these special cases, the
methodology is limited to controlling only those harmonics of the
$H(\phi)$ which do not correspond to the vanishing harmonics in
$Z(\phi)$.

It is straightforward to extend the above arguments to a population of
oscillators. Under the assumption that a parameter in each individual oscillator
can be independently tuned online, we may consider
\begin{equation}
  \frac{d \bm{A}_i (t)}{dt} = \bm{F}_i [\bm{A}_i(t)]  + K (0,p_i,\ldots)^T,
   \label{full_equation2}
\end{equation}
where the function $p_i$ is fully generalized as
\begin{equation}
 p_i(t) = \sum_{j} \sum_{n=0}^S  k_{ij}^{(n)} y_j^n (t-\tau_{ij}^{(n)}).
  \label{general_p}
\end{equation}
The corresponding phase model then reads Eq. (\ref{pm_general}).
The interaction function $H_{ij}(\phi)$ is determined as a function of the
physical feedback parameters $\{k_{ij}^{(n)}\}$ and
$\{\tau_{ij}^{(n)}\}$; Any $H_{ij}(\phi)$ can be designed by giving
appropriate feedback parameters.  The phase model (\ref{pm_general}) is
very general and a large class of collective behavior can be engineered.

A simple situation, which is the case in the numerical and
experimental studies in Secs. \ref{sec:clustering} and
\ref{sec:experiments}, is that a global parameter of the system is tuned
by feedback. In such a case, we consider the phase model (\ref{pm}) by
adopting the global feedback signal (\ref{p}) and (\ref{h}).

\section{Use of Harmonic Signals} \label{sec:harmonic}
An oscillator may appear to have a nearly harmonic waveform by the
nature of oscillation or through the use of a lowpass filter.  In the
case of a perfectly harmonic waveform, feedback parameters can be
explicitly calculated and the effect of each feedback term on the
interaction function clearly understood.  However, it is unrealistic
to assume that an exact harmonic waveform can be obtained
experimentally.  Therefore, the effect of a weakly inharmonic waveform
is also examined.

\subsection{Harmonic Waveform} \label{sec:exactly_harmonic}
We first assume that $x(\phi)$ is an exact harmonic waveform with zero mean.
Properly defining the origin of the phase, we may set
\begin{equation}
 x(\phi)= e^{-i\phi} + {\rm c.c.}
 \label{xcccgr}
\end{equation}
where c.c. indicates the complex conjugate (i.e., $e^{i \phi}$ in this
case). 

Introducing $\phi_n \equiv \phi-\omega \tau_n$, we obtain
\begin{equation}
 x(\phi_n)^n =\sum_{m=0}^{n} C^{n}_{m} e^{i(n-2m)\phi_n}
\end{equation}
where $C^n_m$ is a number of $m$ combinations from a set of $n$
elements.  Since $x(\phi_n)^n$ contains only $l^{\rm th}$ harmonics (where
$l=n,n-2,n-4,\ldots$) the $n^{\rm th}$ feedback term in
Eq.~(\ref{convolution}) only contributes to the $l^{\rm th}$
($l=n,n-2,n-4,\ldots$) Fourier components of the interaction function.
The feedback delay varies the ratio between the even and odd components
of each harmonic.

Combining the feedback terms using Eq. (\ref{h}) yields an interaction
function composed of Fourier components $H_l$ for $l\leq S$.
\begin{eqnarray}
 h(\phi) =\sum_{n=0}^S \sum_{m=0}^{n} C^{n}_{m} e^{i(n-2m) \phi_n}
\end{eqnarray}
or, for $l \ge 0$
\begin{eqnarray}
 h_l = \sum_{n=l}^S k_n C^n_{\frac{n+l}{2}} e^{il \omega \tau_n}
\end{eqnarray}
where for convenience we define $C^n_m=0$ if $m$ is
not an integer. Therefore
\begin{equation}
 H(\phi) = \sum_{n=0}^S k_n \sum_{m=0}^{n} C^n_m z_{n-2m} e^{i (2m-n)\phi_n},
  \label{H_harmonic}
\end{equation}
or, for $ l \ge 0$
\begin{equation}
 H_l = z_{-l} \sum_{n=l}^S k_n  C^n_{\frac{n+l}{2}} e^{i l \omega \tau_n},
  \label{h_harmonic}
\end{equation}
Therefore, for a given target $H_l$ ($l=0,\ldots,S$), the feedback
parameters $k_n$ and $\tau_n$ ($n=0,\ldots,S$) can explicitly obtained.
Since $H_S$ is determined solely by the $S^{\rm th}$ term, the
parameters $k_S$ and $\tau_S$ can be found by solving a complex equation
obtained by setting $l=S$.  The same process can be used for the
$H_{S-1}$ component, to solve for the parameters $k_{S-1}$ and
$\tau_{S-1}$.  Since $H_{S-2}$ is dependent on the $S^{\rm th}$ and
$(S-2)^{th}$ terms, $k_{S-2}$ and $\tau_{S-2}$ can be determined.
Repeating this processes for each term, all feedback parameters can be
calculated.

As an illustration, the feedback parameters required to produce a
Hansel-Manubille-Mato type \cite{hansel93} interaction function will be
calculated.  The target function $H(\phi)$ is
selected to be
\begin{equation}
 H(\phi) = \sin(\phi-\alpha) - r \sin (2\phi)
 = -\frac{i}{2}e^{i\alpha}e^{-i\phi} + 
  \frac{ir}{2}e^{-2i\phi} + {\rm c.c.},
  \label{hansel}
\end{equation}
where $\alpha$ and $r>0$ are the parameters of the function.
Since the target interaction function has second order components, second order feedback 
is required.  Therefore, equation (\ref{H_harmonic}) for $S=2$ becomes
\begin{equation}
 H(\phi) = (k_0 + 2k_2)z_0 + k_1 z_{-1} e^{i\omega\tau_1}e^{-i\phi}
  + k_2 z_{-2}  e^{2i\omega\tau_2} e^{-2i\phi} + {\rm c.c.}.
  \label{H_second}
\end{equation}
Comparing Eq.~(\ref{hansel}) and Eq.~(\ref{H_second}), we find one of
the solutions to be
\begin{equation}
 k_0 = - \frac{r}{|z_{-2}|},\quad
 k_1 = \frac{1}{2 |z_{-1}|} ,\quad
 k_2 = \frac{r}{2 |z_{-2}|}, 
\end{equation}
\begin{equation}
 \tau_1 = \frac{\alpha-\frac{\pi}{2}-\arg(z_{-1})}{\omega},\quad
 \tau_2 = \frac{\frac{\pi}{2}-\arg(z_{-2})}{2\omega}.
\end{equation}

\subsection{Slightly Inharmonic Waveform} \label{sec:inharmonic}
The effect of weak inharmonic components is considered using the waveform
\begin{equation}
  x(\phi)= e^{i\phi} + \epsilon e^{i2\phi}  + O(\epsilon^2)+ \rm{c.c.},
   \label{inharmonic}
\end{equation}
where $\epsilon$ is a small complex number. Introducing
$\phi_n=\phi-\omega \tau_n$, for $n\ge 1$ yields
\begin{equation}
 x(\phi_n)^n = \sum_{m=0}^{n} C^{n}_{m} e^{i(n-2m)\phi_n} +
  \epsilon \sum_{m=0}^{n-1} C^{n-1}_{m} \left( e^{i(n-2m+1) \phi_n} +
					 e^{i(n-2m-3) \phi_n}\right) +
  O(\epsilon^2), 
  \label{cgreqn2}
\end{equation}
and $x^0 = 1$.  The $n^{\rm th}$ term
contributes to the $l^{\rm th}$ ($l=n+1,n-1,\ldots$) harmonic with order
$\epsilon$.  $S^{\rm th}$ order feedback strongly enhances the
harmonics $h_l$ of the interaction function for $l \leq
S$. Therefore, $S^{\rm th}$ order feedback is required to produce a target
interaction function composed of harmonics $l \leq S$.  Although the
$(S+1)^{\rm th}$ Fourier component appears in the interaction function, it
is expected to be very small [of the order of $O(\epsilon Z_{S+1})$] and can
be safely neglected. Similarly, this result is also true in
cases where the first order Fourier component of the waveform is dominant.

When the waveform is strongly inharmonic, each feedback term enhances
various harmonics, including higher order harmonics $H_l$ $(l>S)$ of the
interaction function. In these situations, no analytical solution is
possible, and the feedback parameters must be numerically optimized
using Eq.~(\ref{H}) or Eq.~(\ref{convolution}). In addition, usually, a
high order feedback is required (i.e. large $S$) such that $Z_l$ for
$l>S$ are negligible.

\section{Numerical Study: Phase Clustering}  \label{sec:clustering}
\begin{figure}
 \includegraphics[width=8.5cm]{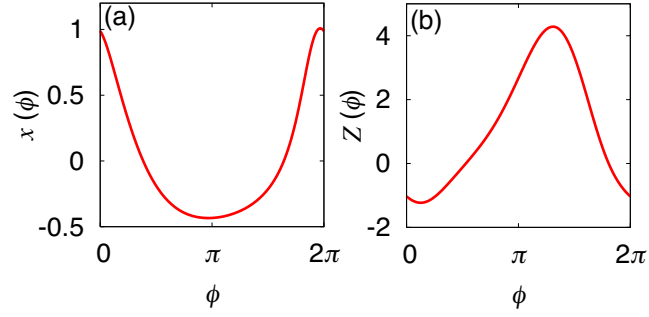}
 \caption{(Waveform $x(\phi)$ and phase response function $Z(\phi)$ of
 the Brusselator oscillator. The parameter values are $A=1.0$ and
 $B=2.3$, resulting in $T=6.43$ and $\omega=0.977$.}  \label{fig:wave}
\end{figure}

Our methodology is numerically verified for a population of limit-cycle
oscillators, using the Brusselator model, a simple two variable ODE
system that admits a Hopf bifurcation \cite{glansdorff71}.  The
dynamical equations for a Brusselator population under global feedback
are
\begin{eqnarray}
 \frac{dx_i}{dt} &=& (B-1) x_i + A^2 x_i + f(x_i,y_i) +
  \frac{K}{N}\sum_{j=1}^N h(x_j) , \nonumber \\
 \frac{dy_i}{dt} &=& -B x_i - A^2 y_i -f(x_i,y_i) 
 \label{bl}
\end{eqnarray}
where $f(x,y) = \frac{B}{A} x^2 + 2Axy + x^2 y$. Here, it is assumed
that the feedback signal is constructed from and applied to the
variables $x_i$.  Note that for convenience, the variables $x_i$ and
$y_i$ are transformed such that the fixed point is shifted to
$(x,y)=(0,0)$.  For a single uncoupled oscillator, the Hopf bifurcation
occurs at $B = B_{\rm c} \equiv 1+A^2$. The parameters of Eq.~(\ref{bl})
were chosen to be $A=1.0$ (so that $B_{\rm c}=2.0$) and $B=2.3$.  The
waveform $x(\phi)$ and the response functions $Z(\phi)$ along the
$x$-direction \footnote{The response function is calculated using
xppaut.} are displayed in Fig. \ref{fig:wave}, and their Fourier
coefficients can be found in Table \ref{tab:fourier}.
\begin{table}
  \begin{tabular}{c|c|c|c|c|c|c}
    $l$  & 0 & 1&2&3&4&5\\ \hline \hline
   Re $a_l$ &  0.00 & 0.30& 0.12& 0.05& 0.02 & 0.00\\ 
   Im $a_l$ & -- & $0.00$ & $0.00$ & $-0.01$ & $-0.01$ & $-0.01$ \\
    $|a_l|$ & 0.00 & 0.30 & 0.12 & 0.05 & 0.02 & 0.01   \\ \hline
   Re $Z_l$ & $1.19$ & $-0.98$ & $-0.20$ & $0.04$ & $0.01$ & $0.00$\\ 
   Im $Z_l$ & -- & $-0.85$ & $0.21$ & $0.01$ & $ - 0.01$ & $-0.00$\\
   $|Z_l|$  & 1.19 & 1.29 & 0.28 & 0.04 & 0.01 & 0.00   \\ 
  \end{tabular}
\caption{Brusselator ($A=1, B=2.3$). Fourier coefficients of the wave
 form $x(\phi)$ and the phase response function
 $Z(\phi)$.}\label{tab:fourier}
\end{table}

The phase of an oscillator $\phi_i$ is determined by measuring the times at which
the orbit in $(x_i,y_i)$ state space crosses a particular point on its limit cycle. 
The phase of the oscillator $\phi_i(t)$ for $t_{m-1} \le t < t_m$ can be defined as
\begin{equation}
 \phi_i(t) = \frac{2 \pi (t-t_{m-1})}{t_{m}-t_{m-1}}.
\end{equation} 
where $t_m$ is defined as the time at which the $m^{\rm th}$ crossing occurs.  
Note that $0 \le \phi_i(t) < 2\pi$.
To quantify the amount of collective global order within the system, it is useful to introduce the
order parameter $R_k$:
\begin{equation}
 R_k = \left| \frac{1}{N}\sum_{j=1}^N e^{i k \phi_j}  \right|.
\end{equation}
For a large population (i.e. large $N$), $R_k=0$ for uniform phase
distribution and $R_k=1$ for a {\rm balanced $k$ cluster
state}, in which the population is split into $k$ equally
populated point clusters distributed uniformly in phase (see Appendix \ref{sec:cluster_state}).

Phase clustering commonly appears in globally coupled oscillator systems
\cite{golomb92,okuda93}.  In systems of identical coupled oscillators,
these states always exist independent of the interaction function, such
that the only outstanding issues to be addressed are the stability and
the basin of attraction of the states.  In this example, four parameter
sets are created [one for each cluster state ($n=1,2,3,4$)] with the
following conditions:
\begin{itemize}
 \item[(i)] the $n$ cluster state is uniquely stable among the balanced
	    cluster states
 \item[(ii)] the cluster state has high linear
	    stability
 \item[(iii)] small amplitude feedback is preferable.
\end{itemize}
For condition (i), it is convenient to use a target interaction function
of the form ${\rm Im} H_n>0$ and ${\rm Im} H_l \leq 0$ for $l \neq n$
(note that the symmetric parts ${\rm Re} H_l$ are irrelevant to the
stability of the balanced cluster states, so that we may arbitrarily set
the symmetric parts). For such an interaction function, the maximum
eigenvalue is given by $\lambda_{\rm max}^{(n)} = - \sum_{l=1}^{\infty}
2 l~{\rm Im} H_{nl}$ (see Appendix \ref{sec:cluster_state} for the
details). We thus require $\lambda_{\rm max}^{(n)}<0$.
Satisfying condition (ii) requires that ${\rm Im} H_n$ is large enough
for high stability.  To satisfy condition (iii),
$n^{\rm th}$ order feedback is used to generate the $n^{\rm th}$ cluster state
(i.e. $S=n$), since the $n^{\rm th}$ cluster state requires $n^{\rm th}$ order
harmonics in its interaction function.
  

\begin{table}
  \begin{tabular}{c|c||c|c||c|c||c|c||c|c}
    \multicolumn{2}{c||}{$n$}  & \multicolumn{2}{c||}{1} &
   \multicolumn{2}{c||}{2} & \multicolumn{2}{c||}{3} & \multicolumn{2}{c}{4} \\ \hline\hline
   \multicolumn{2}{c||}{Im~$H_1$}& $> 1.0$ & $1.00$ & $<-1.0$ & $-1.00$&
   $<-1.0$& $-3.28$ & $<-1.0$& $-9.04$ \\ 
   \multicolumn{2}{c||}{Im~$H_2$}& $< 0.0$ & $-0.07$& $>0.3$  & $0.30$ &
   $<-0.4$& $-0.40$ & $<-0.4$& $-3.27$ \\ 
   \multicolumn{2}{c||}{Im~$H_3$}& $< 0.0$ & $-0.01$& $<0.0$  & $-0.00$&
   $>0.2$& $0.20$ & $<-0.2$ & $-0.20$\\
   \multicolumn{2}{c||}{Im~$H_4$}& $< 0.0$ & $-0.00$& $<0.0$  & $-0.00$&
   $<0.0$ &$-0.02$ & $>0.15$ & $0.15$ \\ \hline
   $k_1$ & $\tau_1$
   & $-2.56$ & $2.40$ & $2.01$ & $2.06$ & $0$ & $0$ &$0$ & $0$ \\ 
   $k_2$ & $\tau_2$
   & --      & --     & $-6.50$ & $0.44$ & $35.7$ & $2.95$ &$0.25$ &$5.26$\\ 
   $k_3$ & $\tau_3$
   & --      & --     & --      & --    & $19.3$ & $0.68$ & $68.6$ &$3.61$\\
   $k_4$ & $\tau_4$
   & -- & --      & --     & --      & --  & -- & $42.0$ &$0.32$ \\ \hline
   \multicolumn{2}{c||}{$\lambda_{\rm max}^{(n)}$} &\multicolumn{2}{c||}{$-1.72$}&
   \multicolumn{2}{c||}{$-1.16$}&\multicolumn{2}{c||}{$-1.18$}&\multicolumn{2}{c}{$-0.80$}   \\
  \end{tabular}
\caption{Brusselator population with global feedback. The target and
 resulting interaction functions, feedback parameters, and the resulting
 maximum eigenvalue for the $n$ cluster state.}\label{tab:cluster}
\end{table}
\begin{figure}
 \includegraphics[width=12cm]{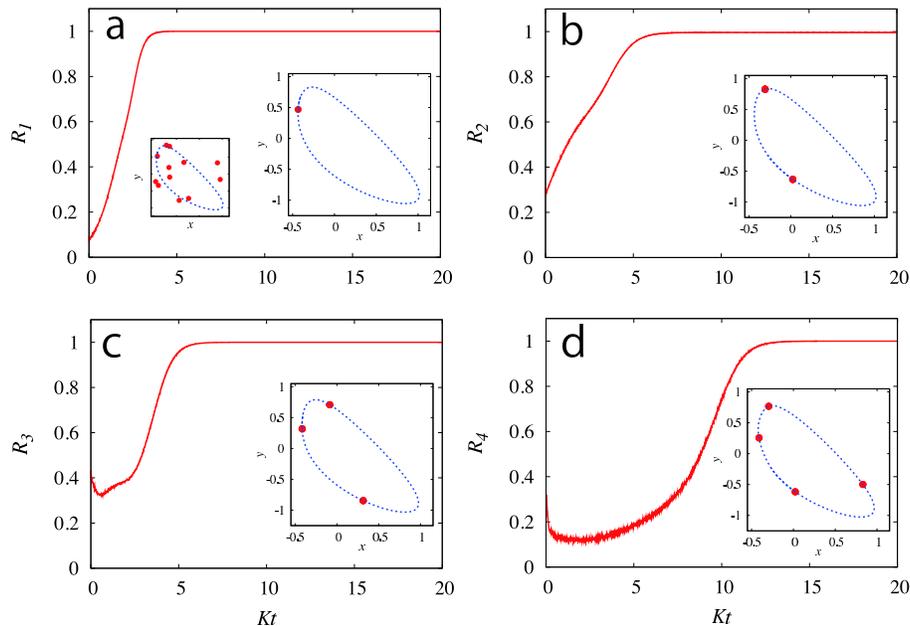} \caption{(Color online) Engineering
 cluster states in the Brusselator model. Time traces of the order
 parameters $R_n$ and snapshots for (a)$n=1$, (b)$n=2$, (c) $n=3$ and
 (d)$n=4$ for $N=12$ and $K=0.001$. In each panel, the parameter set $n$
 diplayed in Table \ref{tab:cluster} is used. In the insets, snapsh ots
 in the one-oscillator phase space at $Kt=20$ are displayed. The dashed
 lines are the orbits of an oscillator. In panel (a), the initial
 conditions are shown (on the left side). Note that the initial
 condition is the same for all panels.  }  \label{fig:order}
\end{figure}

A large number of interaction functions satisfies conditions (i) and
(ii).  Out of this family of interaction functions, the optimal feedback
parameter set is selected such that it minimizes the cost function
$\sum_{l=1}^{n} |k_l|$ under the conditions shown in the left side of
each column of Table \ref{tab:cluster}.  The table also displays the
optimized feedback parameter sets (obtained numerically using
Mathematica), the resulting ${\rm Im} H_l$ (in the right side of each
column), and the resulting maximum eigenvalue $\lambda_{\rm max}^{(n)}$.

Applying the optimized feedback parameter sets to Eq. (\ref{bl}) causes
the system to approach the desired cluster states.  The convergence of
the system to the cluster states is illustrated in Fig. \ref{fig:order},
using the appropriate order parameter $R_n$.  Several different random
initial conditions were used for each parameter set and in each case the
desired cluster state was obtained (not shown).

It is worth noting that, in practice, the 3 and 4 cluster
states are difficult to obtain unless high order
feedback is used. For example, when only the linear term is used,
the magnitude $|H_l|=|Z_l a_l|$ is very small for $l \ge
3$. This fact implies that the maximum eigenvalue of the $n \ge 3$
cluster states can not be large and negative. Therefore, the presence of
noise or heterogeneity, if any, would destroy the $n \ge 3$
cluster states.

\section{Experimental Studies} \label{sec:experiments}
\subsection{Experimental Setup} \label{sec:ExpClustering}
The preceding theoretical work on synchronization engineering was
experimentally tested using a population of electrochemical oscillators.
These oscillators were created using an electrochemical cell which
consisted of 64 Ni electrodes (99.99\% pure) in a 3M H$_2$SO$_4$
solution, a Pt mesh counter electrode, and a Hg/Hg$_2$SO$_4$/K$_2$SO$_4$
(sat) reference electrode.  The cell was enclosed in a jacketed glass
vessel held at a constant temperature of $11^{\rm o}$C.  An EG\&G
potentiostat was used to adjust the circuit potential $(V)$ of the cell,
causing the nickel electrodes to undergo transpassive dissolution.  The
dissolution current of each electrode, $I_j(t)$, was measured by zero
resistance ammeters (ZRA).  A resistor ($R_p$) was attached to each
channel to induce oscillations in the electrode potential.  A Labview
based real time data acquisition computer was used to read the ZRA
measurements, stream these measurements to the host machine, and apply
the feedback signal to the potentiostat at a rate of 250 Hz.  The
current measurements were scaled:
\begin{equation}
 I'_{j}(t) = \frac{A_{\rm mean}}{A_j}(I_j(t) - I_j^{\rm offset})
  \label{CGR1}
\end{equation}       
The mean value of each channel ($I_j^{\rm offset}$) was removed from the
measurement, and the result was scaled by the amplitude of its
oscillation ($A_j$) relative to the mean amplitude of the population
($A_{\rm mean}$).  The host machine was used to continuously determine the
offset and amplitude of each rhythmic element in the population.  To
calculate the feedback signal, the potential drop across the double
layer, $x_j(t)$, was determined from the scaled current measurements
\begin{equation}
	x_j(t) = V(t) - I'_{j}(t)R_p
	\label{CGR2}
\end{equation}
where $V(t)$ is the applied voltage.  The perturbation signal $p(t)$
which was fed back to the potentiostat, $V(t)=V_0+Kp(t)$, was calculated
by taking the mean value of $h(x_j(t))$ over every element in the
population,
\begin{equation}
	p(t) = \frac{1}{N} \sum_{j=0}^{N} h(x_j(t))
	\label{CGR3}
\end{equation}        
\begin{equation}
	h(x(t)) = \sum_{n=0}^{S} k_n x(t-\tau_n)^n
	\label{CGR4}
\end{equation}                                                       
where $K$ is the overall feedback gain, $N$ is the number of elements in
the population, $k_n$ is the $n^{\rm th}$ polynomial feedback coefficient,
$\tau_n$ is the time delay of the nth polynomial feedback term, and $S$
is the polynomial feedback order.

\subsection{Experimental Validation of Theory} \label{sec:ExpTheoryValid}
\begin{figure}[t]
 \centering \includegraphics[width=17cm]{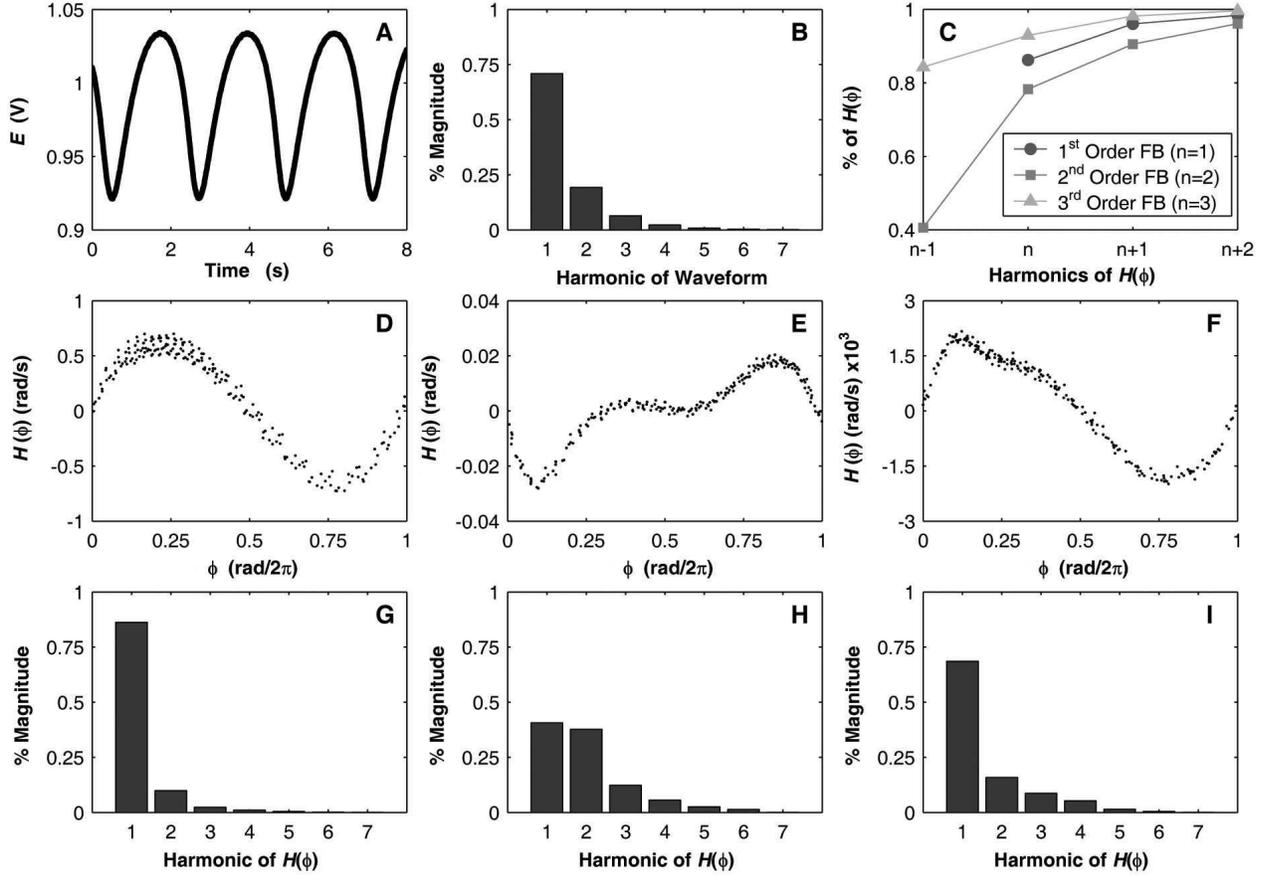} \caption{Electrochemical
 experiments. (A) Time series of a single nearly-smooth oscillator
 ($V_0=1.110$ V, $R_p=650$ $\Omega$).  (B) Percentage of the first seven
 harmonics within the waveform of the rhythmic element.  (C) The
 percentage of $H(\phi)$ contained within the first $k$ harmonics of
 $H(\phi)$ as a function of the choice of $k$.  The values of $k$ have
 been re-centered by the feedback order $n$.  (D-F) Experimental
 measurements of interaction functions corresponding to (D) $1^{\rm st}$
 order feedback ($K=0.07$, $k_0=0$ V, $k_1=1$, $\tau_1=0.013$
 rad/$2\pi$), (E) $2^{\rm nd}$ order feedback ($K=1.6$, $k_0=-0.003$ V,
 $k_1=0$, $k_2=1$ V$^{-1}$, $\tau_2=0.013$ rad$/2\pi$), and (F) $3^{\rm
 rd}$ order feedback ($K=18$, $k_0=6.5\times10^{-5}$ V, $k_1=0$, $k_2=0$
 V$^{-1}$, $k_3=1$ V$^{-2}$, $\tau_3=0.013$ rad$/2\pi$).  (G-I)
 Percentage of the first 7 harmonic components within the measured
 interaction function using (G) $1^{\rm st}$ order feedback, (H) $2^{\rm
 nd}$ order feedback, and (I) $3^{\rm rd}$ order feedback.}
 \label{CGR_FIG_H_Enhancement}
\end{figure}

The synchronization engineering framework, as derived in Sec.
\ref{sec:inharmonic} for weakly inharmonic waveforms, predicts that
$n^{\rm th}$ order feedback can enhance up to and including the $n^{\rm
th}$ harmonics of the interaction function for harmonic waveforms, and
$(n+1)^{\rm th}$ harmonics for weakly inharmonic waveforms.  To test the
range of the validity of this result, the experimental system was used
to measure how the harmonics of an interaction function change with
increasing feedback order.  The operating voltage of the system was
selected to be 1.110 V as this was found to be close enough to the Hopf
bifurcation to produce a nearly harmonic waveform, but far enough to
ensure that the periodic cycle was robust against external perturbations
[Figs. \ref{CGR_FIG_H_Enhancement}(A) and \ref{CGR_FIG_H_Enhancement}(B)].

A two oscillator system was used to measure the interaction function
associated with the global feedback signal.  This method of measurement
was created by extending the work of Miyazaki and Kinoshita
\cite{Miyazaki06} to rhythmic systems under global feedback.  By
measuring the change in the period of the two elements as a function of
their phase difference, the interaction function can be experimentally
determined.

While the $1^{\rm st}$ order harmonic of the waveform accounts for
$~71\%$ of its magnitude, this may not be sufficient to allow the
$O(\epsilon)$ terms of Eq. (\ref{cgreqn2}) to be neglected.  Therefore,
it is expected that the $(n+1)$ order harmonics of the interaction
function will be dominant.  Figure \ref{CGR_FIG_H_Enhancement}(C)
illustrates the percentage of the cumulative magnitude of the harmonics
of $H(\phi)$ as a function of the choice of the highest harmonic
component to be considered.  The cutoff harmonic is given in terms of
the feedback order to allow different orders of feedback to be compared
to one another.  Applying a $1^{\rm st}$ order feedback signal to the
experimental system produced an interaction function with a large
$1^{\rm st}$ order component and a relatively small $2^{\rm nd}$ order
component [Figs. \ref{CGR_FIG_H_Enhancement}(D) and
\ref{CGR_FIG_H_Enhancement}(G)].  While the $1^{\rm st}$ order harmonic
only makes up $~82\%$ of $H(\phi)$, the combination of the $1^{\rm st}$
and $2^{\rm nd}$ order harmonics account for $~96\%$ of its magnitude.
When a $2^{\rm nd}$ order feedback was used, it substantially reduced
the $1^{\rm st}$ order harmonic of $H(\phi)$ while increasing the
$2^{\rm nd}$ order harmonic [Figs. \ref{CGR_FIG_H_Enhancement}(E) and
\ref{CGR_FIG_H_Enhancement}(H)].  A small increase in the $3^{\rm rd}$
order harmonic was observed due to anharmonicity.  Together, these three
components make up $\sim 90\%$ of the overall magnitude of $H(\phi)$.
$3^{\rm rd}$ order feedback increases the ratio between the $3^{\rm rd}$
and $1^{\rm st}$ order harmonics of $H(\phi)$ when compared to $1^{\rm
st}$ order feedback (0.126 for $3^{\rm rd}$ order feedback vs 0.027 for
$1^{\rm st}$ order feedback).  Due to strong second order harmonics in
the waveform, non-trivial $2^{\rm nd}$ and $4^{\rm th}$ order components
were also observed in $H(\phi)$.  In this case, the first four
components account for $\sim 98\%$ of the overall magnitude of the
interaction function.  These results indicate that the overall shape of
the of $H(\phi)$ is largely composed of $l^{\rm th}$ harmonics with $l
\leq n+1$, where $n$ is the feedback order used to produce the function,
in line with theoretical expectations.

\begin{figure}[t]
 \centering \includegraphics[width=17cm]{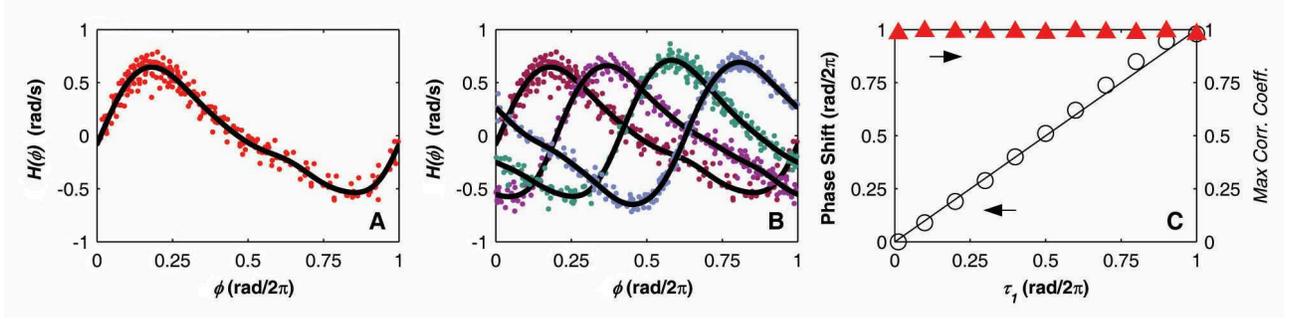} \caption{(Color online)
  (A) Experimental measurements (dots) with a Fourier fit (line) of an
  interaction function obtained using linear feedback ($V_0=1.165$ V,
  $R_p=650$ $\Omega$, $K=0.07$, $k_0=0$ V, $k_1=1$, $\tau_1=0.013$
  rad$/2\pi$).  (B) Interaction functions obtained with
  $\tau_1=[0.013,0.2,0.4,0.6]$ rad$/2\pi$ respectively.  (C) Phase shift
  of the interaction function and maximum value of the correlation
  coefficient as a function of feedback delay. }
  \label{CGR_FIG_PhaseShift}
\end{figure}
While the magnitude of the harmonics of $H$ is controlled by the
feedback order and their associated feedback gains $\{k_n\}$, the ratio
between the symmetric and anti-symmetric components of $H$ is controlled
by the feedback delay $\{\tau_n\}$.  This indicates that increasing the
feedback delay is equivalent to shifting the phase of the corresponding
components of the interaction function.  To validate this claim
experimentally, a series of interaction functions was measured using a
two oscillator system with global first order feedback over a range of
feedback delay $\tau_1$ from 0.013 to 1 rad$/2\pi$.

The base interaction function ($\tau_1=0.013$ rad$/2\pi$) is illustrated
in Fig.  \ref{CGR_FIG_PhaseShift}(A).  As the feedback delay was
increased, the interaction function was observed to shift
[Fig. \ref{CGR_FIG_PhaseShift}(B)].  To determine the phase shift of the
interaction functions when $\tau_1>0.013$ rad$/2\pi$, a correlation
function was calculated between the base interaction function and each
of the shifted interaction functions.  The correlation function was
determined by finding the correlation coefficient between the shifted
interaction functions and the base interaction function as the phase of
the base function was rotated from 0 to $2\pi$.  The phase which
produced the maximum value of the correlation coefficient was taken as
the experimentally observed phase shift.  Figure
\ref{CGR_FIG_PhaseShift}(C) indicates that the phase shift of $H$ is
directly proportional to the feedback delay with a proportionality
constant of 1.  For each measurement, the maximum value of the
correlation coefficients remained high ($>0.98$), indicating that the
overall shape of the interaction function was preserved.

Knowing the relationship between the harmonics of the interaction
function and the feedback parameters, it is possible to engineer a
feedback that produces a desired interaction function, for example,
$H(\phi) = \sin(\phi - \alpha) - r \sin(2\phi)$ where $\alpha=1.32$ and
$r=0.25$.

\begin{figure}
 \centering \includegraphics{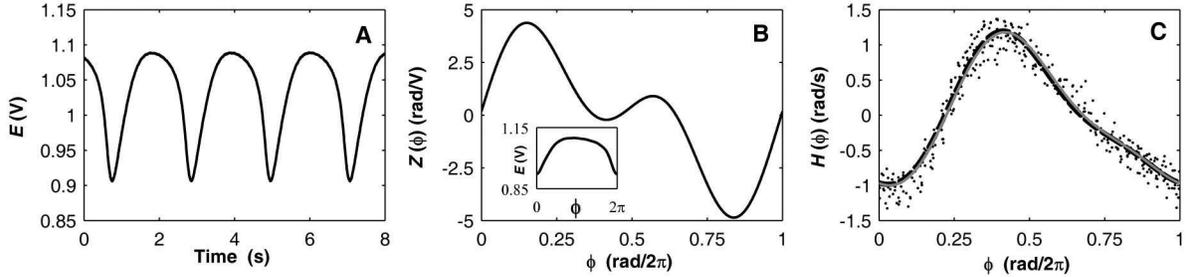} \caption{(A) Time series of electrode
  potential ($V_0=1.165$ V, $R_p=650$ $\Omega$).  (B) Response function
  $Z(\phi)$ and waveform (inset) of a single oscillator.  (C) Target
  (solid line), optimized (dashed line), and measured (dots) interaction
  function with feedback parameters $K=0.0494$, $k_0=-0.0526$ V,
  $k_1=8.7376$, $k_2=16.3696$ V$^{-1}$, $\tau_1=0.21$ rad$/2\pi$,
  $\tau_2=0.68$ rad$/2\pi$.}  \label{CGR_FIG_Engineering_H}
\end{figure}

Before the feedback parameters associated with this interaction function
can be calculated, the waveform [Fig. \ref{CGR_FIG_Engineering_H}(A)]
and the response function of the oscillations must be determined.  The
response function was calculated using Eq. (\ref{H}) from multiple
measurements of interaction functions under different feedback
conditions (usually $1^{\rm st}$, $2^{\rm nd}$ and $3^{\rm rd}$ order
feedback).  Since Eq. (\ref{H}) does not have an analytical solution for
the response function, a numerical optimization algorithm was used to
calculate the Fourier coefficients of $Z(\phi)$
[(Fig. \ref{CGR_FIG_Engineering_H}(B)].  Once the response function was
known, the feedback parameters $\{k_n\}$ and $\{\tau_n\}$ were optimized
to achieve the desired interaction function also using Eq. (\ref{H})
\cite{kiss07}.  The interaction function produced by the optimized
feedback parameters was experimentally determined to ensure that they
produce the expected function.  Figure \ref{CGR_FIG_Engineering_H}(C)
compares the experimentally measured interaction function to the
interaction function predicted by Eq. (\ref{H}).  By calculating the
Fourier coefficients of the experimental measurements, it was determined
that $\alpha=1.350$ and $r=0.242$, within $3\%$ of their target values.

\begin{figure}
 \centering \includegraphics{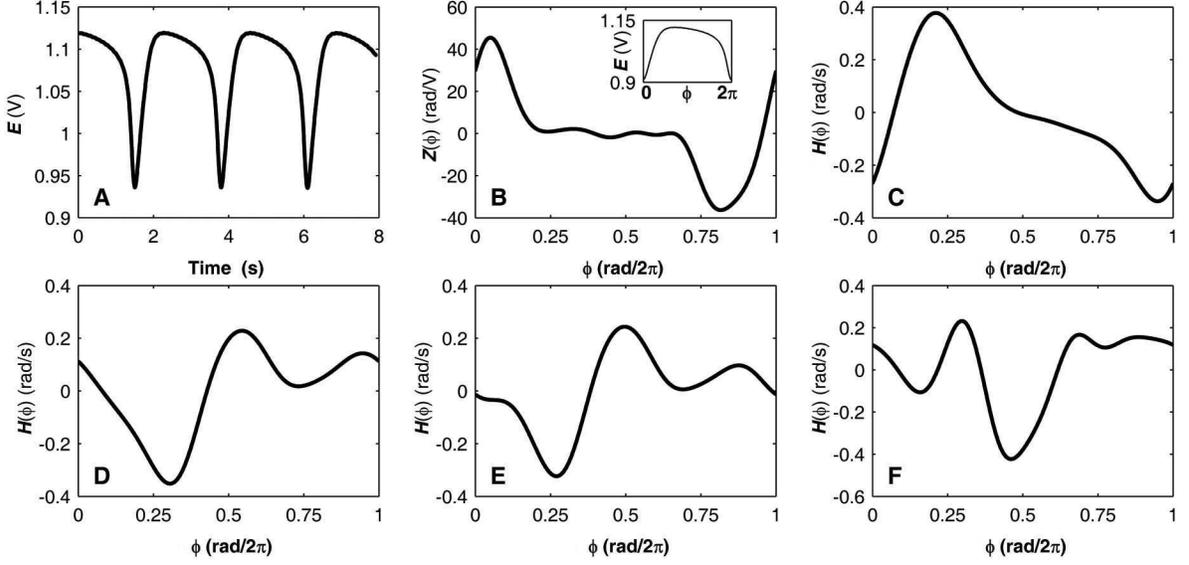} \caption{Experiments: Effects of
  polynomial feedback on the interaction function. (A) Time series of
  electrode potential during electrodissolution of nickel wires in
  sulfuric acid ($V_0=1.195$ V, $R_p=650$ $\Omega$) (B) Response
  function and waveform (inset) of a single oscillator.  (C) Interaction
  function optimized to produce a 1 cluster state.  ($K=0.4$, $k_0=0$ V,
  $k_1=1$, $\tau_1=0.014$) (D) Interaction function optimized to produce
  a 2 cluster state. ($K= 0.0425$, $k_0= 14.97$ V, $k_1= -3.265$, $k_2=
  -66.087$ V$^{-1}$, $\tau_1=0.014$, $\tau_2=0.368$) (E) Interaction
  function optimized to produce a 3 cluster state. ($K= 0.0424$, $k_0=
  20.747$ V, $k_1= -4.142$, $k_2= -72.317$ V$^{-1}$, $k_3= 251.744$
  V$^{-2}$, $\tau_1=0.014$ rad$/2\pi$, $\tau_2=0.32$, $\tau_3=0.04$) (F)
  Interaction function optimized to produce a 4 cluster state.  ($K=
  0.0839$, $k_0= 1.787$ V, $k_1= -5.099$, $k_2= -34.136$ V$^{-1}$, $k_3=
  196.145$ V$^{-2}$, $k_4 = 3139.686$ V$^{-3}$, $\tau_1=0.014$,
  $\tau_2=0.44$, $\tau_3=0.02$, $\tau_4=0.36$).  The feedback delay
  times are given in rad$/2\pi$.  } \label{fig:CGR_FIG_Waveform_and_RF}
\end{figure}
\begin{table}[t]
  \hfill    
  \begin{minipage}{\textwidth}
    \begin{tabular}{c|c|c|c|c|c|c|c|c|c|c|c|c }
     \multicolumn{1}{c}{\;} & \multicolumn{2}{c}{{\scriptsize Waveform}} & \multicolumn{2}{c}{{\scriptsize Response Fn}} & \multicolumn{2}{c}{$H$ {\scriptsize(1 Cluster)}} & \multicolumn{2}{c}{$H$ {\scriptsize(2 Cluster)}} & \multicolumn{2}{c}{$H$ {\scriptsize(3 Cluster)}} & \multicolumn{2}{c}{$H$ {\scriptsize(4 Cluster)}}\\
     \hline
     $n$ & \multicolumn{2}{c|}{{\scriptsize Even \ \;\; Odd \ }} & \multicolumn{2}{c|}{{\scriptsize Even \ \;\; Odd \ }} & \multicolumn{2}{c|}{{\scriptsize Even \ \;\; Odd \ }} & \multicolumn{2}{c|}{{\scriptsize Even \ \;\; Odd \ }} & \multicolumn{2}{c|}{{\scriptsize Even \ \;\; Odd \ }} & \multicolumn{2}{c}{{\scriptsize Even \  \;\; Odd} \ } \\
     \hline 
     \;1\; & {\scriptsize-0.0710} & {\scriptsize+0.0063} & {\scriptsize+5.6533} & {\scriptsize+15.460} & {\scriptsize-0.0799} & {\scriptsize+0.2211} & {\scriptsize-0.0014} & {\scriptsize-0.1688} & {\scriptsize-0.0619} & {\scriptsize-0.1289} & {\scriptsize+0.1503} & {\scriptsize-0.0617} \\
     2 & {\scriptsize-0.0399} & {\scriptsize-0.0056} & {\scriptsize+12.229} & {\scriptsize+14.496} & {\scriptsize-0.1303} & {\scriptsize+0.0819} & {\scriptsize+0.1475} & {\scriptsize+0.0446} & {\scriptsize+0.1313} & {\scriptsize-0.0157} & {\scriptsize-0.1384} & {\scriptsize-0.0291} \\
     3 & {\scriptsize-0.0212} & {\scriptsize-0.0054} & {\scriptsize+9.9103} & {\scriptsize+5.7058} & {\scriptsize-0.0507} & {\scriptsize+0.0010} & {\scriptsize-0.0473} & {\scriptsize-0.0138} & {\scriptsize-0.0657} & {\scriptsize+0.0353} & {\scriptsize+0.1281} & {\scriptsize-0.0593} \\
     4 & {\scriptsize-0.0133} & {\scriptsize-0.0026} & {\scriptsize+1.8530} & {\scriptsize+2.6015} & {\scriptsize-0.0071} & {\scriptsize+0.0029} & {\scriptsize-0.0101} & {\scriptsize-0.0127} & {\scriptsize-0.0140} & {\scriptsize-0.0148} & {\scriptsize-0.0049} & {\scriptsize+0.0414} \\
     5 & {\scriptsize-0.0064} & {\scriptsize-0.0005} & {\scriptsize-0.8616} & {\scriptsize+2.1364} & {\scriptsize-0.0004} & {\scriptsize+0.0031} & {\scriptsize-0.0067} & {\scriptsize+0.0013} & {\scriptsize-0.0009} & {\scriptsize-0.0057} & {\scriptsize-0.0275} & {\scriptsize+0.0069} \\
     6 & {\scriptsize-0.0038} & {\scriptsize+0.0004} & {\scriptsize+0.5878} & {\scriptsize+1.7258} & {\scriptsize-0.0010} & {\scriptsize+0.0011} & {\scriptsize-0.0033} & {\scriptsize-0.0010} & {\scriptsize-0.0021} & {\scriptsize+0.0024} & {\scriptsize+0.0130} & {\scriptsize-0.0022} \\    
    \hline           
    \end{tabular}    
    \vspace{15pt}
  \end{minipage} 
   
  \begin{minipage}{\textwidth}
   \begin{minipage}[t]{.47\textwidth}
    \begin{tabular}{c c c c c}
    \hline
        $1 \text{ cluster}$ & $\lambda_1$ & $\lambda_2$ & $\lambda_3$ & $\lambda_4$ \\
    \hline
        $M = 1$ & {\scriptsize $-0.422$} & $ -  $ & $ - $ & $ - $ \\
        $M = 2$ & {\scriptsize $+0.058$} & {\scriptsize $-0.182$} & $ - $ & $ - $ \\
        $M = 3$ & {\scriptsize $+0.196$} & {\scriptsize $+0.196 $} & {\scriptsize $-0.010$} & $ - $ \\
        $M = 4$ & {\scriptsize $+0.108 $} & {\scriptsize $+0.159 $} & {\scriptsize $+0.108 $} & {\scriptsize $-0.012$} \\
    \hline
    \\
    \hline
        $3 \text{ cluster}$ & $\lambda_1$ & $\lambda_2$ & $\lambda_3$ & $\lambda_4$ \\
    \hline
        $M = 1$ & {\scriptsize $+0.127 $} & $ -  $ & $ - $ & $ - $ \\
        $M = 2$ & {\scriptsize $+0.025 $} & {\scriptsize $+0.076 $} & $ - $ & $ - $ \\
        $M = 3$ & {\scriptsize $-0.245$} & {\scriptsize $-0.245$} & {\scriptsize $-0.121$} & $ - $ \\
        $M = 4$ & {\scriptsize $+0.034 $} & {\scriptsize $+0.043 $} & {\scriptsize $+0.034 $} & {\scriptsize $+0.059$} \\
    \hline
   \end{tabular}
  \end{minipage}
  \hfill
  \begin{minipage}[t]{.47\textwidth}
   \begin{tabular}{c c c c c}
    \hline
        $2 \text{ cluster}$ & $\lambda_1$ & $\lambda_2$ & $\lambda_3$ & $\lambda_4$ \\
    \hline
        $M = 1$ & {\scriptsize $+0.172$} & $ -  $ & $ - $ & $ - $ \\
        $M = 2$ & {\scriptsize $-0.236$} & {\scriptsize $-0.032$} & $ - $ & $ - $ \\
        $M = 3$ & {\scriptsize $-0.014$} & {\scriptsize $-0.014$} & {\scriptsize $+0.048 $} & $ - $ \\
        $M = 4$ & {\scriptsize $-0.051$} & {\scriptsize $+0.134 $} & {\scriptsize $-0.051$} & {\scriptsize $+0.051 $} \\
    \hline
    \\
    \hline
        $4 \text{ cluster}$ & $\lambda_1$ & $\lambda_2$ & $\lambda_3$ & $\lambda_4$ \\
    \hline
        $M = 1$ & {\scriptsize +0.111} & $ -  $ & $ - $ & $ - $ \\
        $M = 2$ & {\scriptsize -0.299} & {\scriptsize -0.094} & $ - $ & $ - $ \\
        $M = 3$ & {\scriptsize +0.231} & {\scriptsize +0.231} & {\scriptsize +0.191} & $ - $ \\
        $M = 4$ & {\scriptsize -0.268} & {\scriptsize -0.237} & {\scriptsize -0.268} & {\scriptsize -0.166} \\
    \hline

    \end{tabular}
   \end{minipage}
  \end{minipage}
  \hfill \caption{(Top) Fourier coefficients of the waveform, response
  function, and the optimized interaction functions for each of the four
  experimental objectives.  (Bottom) Transversal eigenvalues for cluster
  states 1--4 for each of the four experiments, as calculated from the
  Fourier coefficients of the corresponding interaction function. }
  \label{fig:CGR_TABLE_FFT_Data}
\end{table}

\subsection{Phase Clustering Experiments} \label{sec:ExpPhaseClustering}
To engineer a cluster state in the experimental system, feedback
parameters were selected such that the desired cluster state was
stabilized.  Four sets of experiments were conducted to obtain balanced
cluster states composed of one to four clusters, using a population of
64 oscillators.  Since a four cluster state requires the presence of
fourth order harmonics in the response function, the operation voltage
of the system was set at 1.195 V for each experiment causing weakly
relaxational oscillations [Figs. \ref{fig:CGR_FIG_Waveform_and_RF}(A)
and \ref{fig:CGR_FIG_Waveform_and_RF}(B)]. 
The Fourier coefficients of the waveform and response function 
can be found in Table \ref{fig:CGR_TABLE_FFT_Data}.
 
\begin{table}[t]
 \begin{tabular}{c c c c c c c c}
 \hline Harmonic: & $\;1^{\rm st}\;$ & $\;2^{\rm nd}\;$ & $\;3^{\rm rd}\;$ &
 $\;4^{\rm th}\;$ & $\;5^{\rm th}\;$ & $\;6^{\rm th}\;$ & $\;7^{\rm th}\;$ \\ \hline 2
 Cluster Exp. Lower Bounds (LB) $\;$ & -1.0 & 0.5 & -2.0 & -0.8 & -0.5 &
 -0.50 & -0.50 \\ 2 Cluster Exp. Upper Bounds (UB) $\;$ & -0.5 & 1.0 &
 -0.4 & -0.3 & -0.1 & -0.05 & -0.05 \\ \hline \hline 3 Cluster
 Exp. Lower Bounds (LB) $\;$ & -1.0 & -2.0 & 0.5 & -0.8 & -0.5 & -0.50 &
 -0.50 \\ 3 Cluster Exp. Upper Bounds (UB) $\;$ & -0.5 & -0.4 & 1.0 &
 -0.3 & -0.1 & -0.05 & -0.05 \\ \hline \hline 4 Cluster Exp. Lower
 Bounds (LB) $\;$ & -1.0 & -2.0 & -0.8 & 0.2 & -0.5 & -0.50 & -0.50 \\ 4
 Cluster Exp. Upper Bounds (UB) $\;$ & -0.5 & -0.4 & -0.3 & 0.5 & -0.1 &
 -0.05 & -0.05 \\ \hline
 \end{tabular}
 \caption{Range of the odd Fourier coefficients of $H(\phi)$ used to
 optimize feedback parameters to produce phase cluster states 2--4. }
 \label{exp:FFT_H_Ranges}
\end{table}
 
As seen in the numerical simulations (Sec. \ref{sec:clustering}), there
exists a large number of equally valid target interaction functions
which can produce the desired cluster states.  No specific target
function was selected; The Fourier coefficients of the interaction
function were optimized such that the desired cluster state was uniquely
(or almost uniquely) stabilized.  Given previous numerical results,
$n^{\rm th}$ order feedback was used to produce an $n$ cluster state.
Since linear feedback is sufficient to produce the one cluster state, no
optimization was necessary in this case.  For the higher order cluster
states, a set of penalties were created to describe the fitness of the
interaction function based on the distance between its Fourier
coefficients and an acceptable range of coefficients.  The fitness of
$H(\phi)$ was calculated using the equations
\begin{equation}
  \text{fitness}  =  \text{magnitude} + \text{penalties}, \nonumber
\end{equation}  
\begin{equation}
  \text{magnitude} = K\sum_{n=1}^S \frac{\left|k_n\right|}{10^n}, \nonumber
\end{equation}
\begin{equation}  
  \text{penalties} = \sum_{n=1}^7 P_n,  
\end{equation} 
\begin{equation}
P_n = \left\{
 \begin{array}{cl} \label{alpha}
  \left| B_n - \frac{\text{LB}_n + \text{UB}_n}{2} \right| &
   \mbox{for $B_n < \text{LB}_n$
   or $B_n > \text{UB}_n$},\\
  0 &
   \mbox{for $\text{LB}_n < B_n < \text{UB}_n$}
 \end{array} \right.
 \nonumber
\end{equation}
where $B_i$ is the $i^{\rm th}$ odd Fourier coefficient of $H(\phi)$.
The upper and lower bounds (UB and LB) of the odd Fourier coefficients
were selected such that the desired cluster state would be stable and
the other (up to 6) cluster states in the system would be unstable.  As
previously demonstrated, this requires that the interaction function for
an $n$ cluster state have a large positive $n^{\rm th}$ order harmonic,
and sufficiently negative $m^{\rm th}$ harmonics ($m\neq n$) to
destabilize all other cluster states.  The target Fourier coefficient
ranges reflect this requirement, and are tabulated in Table
\ref{exp:FFT_H_Ranges}.  Additionally, a magnitude adjustment was added
to penalize parameter sets which produced a large amplitude feedback
signal.  Large feedback perturbations are not desirable since they may
change the amplitude of the rhythmic elements of the system, violating
the phase approximation.  By minimizing the value of the fitness
variable, the optimization forced the interaction function to have
Fourier coefficients necessary to produce the desired cluster state.
The optimized interaction functions are illustrated in
Figs. \ref{fig:CGR_FIG_Waveform_and_RF}(C)--\ref{fig:CGR_FIG_Waveform_and_RF}(F).

The transversal eigenvalues of states with one to four clusters can be
seen in Table \ref{fig:CGR_TABLE_FFT_Data} for each experiment.  They
were calculated from the Fourier coefficients of the experimental
interaction functions using Eqs. (\ref{Stab_1}) and (\ref{Stab_2}).
The eigenvalues indicate that the desired one, two, and three cluster
states are uniquely stable.  In the case of the four cluster experiment,
the numerical optimization was unable to find feedback parameters to
stabilize the four cluster state without also stabilizing the two
cluster state.  This is not unexpected, given that the difficulty of the
optimization dramatically increases with feedback order.  Therefore, the
four cluster experiment will have a bi-stability between the four
cluster state and the two cluster state.  In this case, the final state
of the system will be determined by the initial conditions of the
system.

\begin{figure}
 \centering \includegraphics{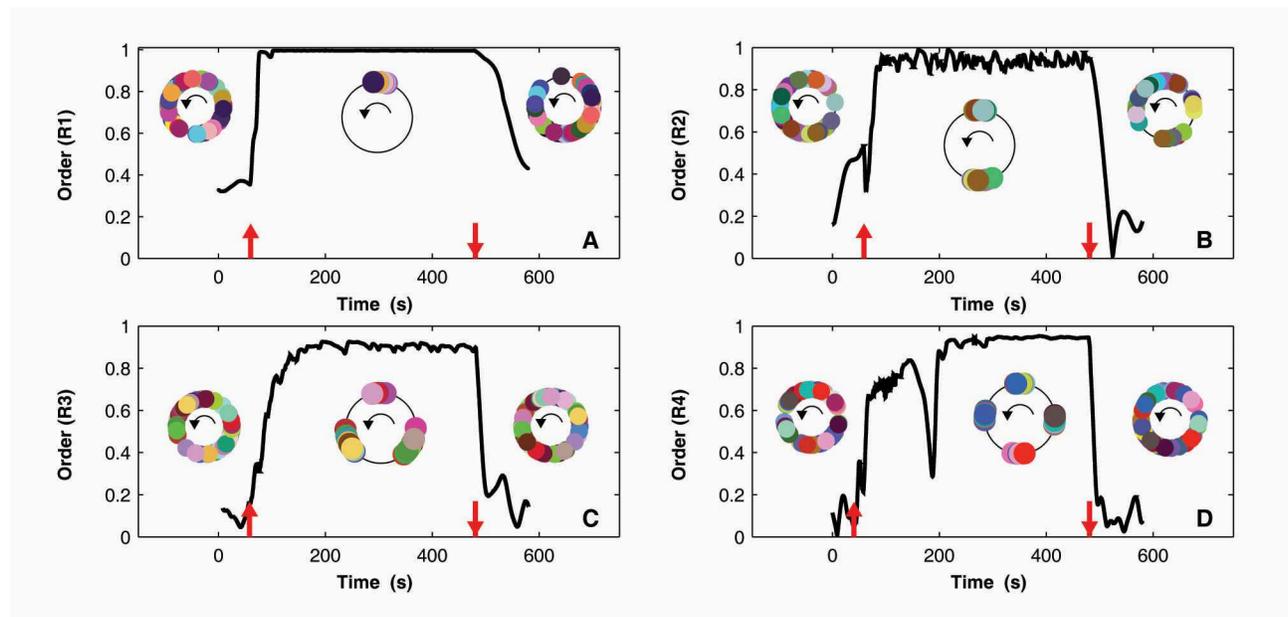} \caption{(Color online) (A) Time
  series of the $R_1$ order parameter, using feedback optimized to
  produce a one cluster state.  Arrows indicate the application and
  termination of the feedback signal. (B) Time series of the $R_2$ order
  parameter using feedback optimized to produce a two cluster state.
  (C) Time series of the $R_3$ order parameter using feedback optimized
  to produce a three cluster state.  (D) Time series of the $R_4$ order
  parameter using feedback optimized to produce a four cluster state.  }
  \label{fig:CGR_FIG_Order_Plot}
\end{figure}

After the feedback parameters were determined, they were applied to the
experimental system, driving it towards the appropriate cluster state
(Fig. \ref{fig:CGR_FIG_Order_Plot}).  Initially no feedback is present in
the system, and the rhythmic elements were isolated from one another.
Without feedback, these elements have a base frequency of 0.5 Hz
$\pm5\%$ with phases randomly distributed between 0 and $2\pi$.  Upon
application of the feedback signal, the system progresses towards the
desired cluster state after a short transient period.  When the feedback
is removed, the system relaxes back to its original unstructured
configuration.  Each experiment was successful in producing the desired
cluster state from the appropriate feedback signal.  It is important to
note that although the four cluster experiment was predicted to have a
bistability between the two and four cluster states, only the four
cluster state was experientially observed.  This seems to indicate that
the basin of attraction for the four cluster state is sufficiently
larger than the basin of the the two cluster state.



\section{concluding remarks}
We have presented a framework for engineering target dynamical behavior
in populations of oscillators with mild feedback. Using a
time delayed, nonlinear feedback, 
Eq.~(\ref{h}), a variety of collective dynamics possible in
weakly coupled oscillators can be engineered. The
comprehensive theory, based on phase models, behind the methodology has
been presented. We have verified the theoretical arguments by both
numerical and experimental studies, showing that the methodology can be
applied accurately to limit-cycle oscillator systems. As an
illustration, by introducing the global feedback given as
Eq. (\ref{p}), various clustering behaviors have been demonstrated
numerically and experimentally.

Our methodology is based on the fact that the existence and stability
conditions of dynamical states in weakly coupled identical oscillators
are characterized by the phase interaction function. Thus, knowing an
interaction function resulting in a target collective dynamics, the only
remaining issue is how to construct the physical interaction yielding
the phase interaction function. An interaction function can be
constructed using the proposed feedback function, Eq. (\ref{h}). The
choice of the specific form of the feedback function was motivated by
the flexible application of the imposed interaction function for
synchronization engineering (Sec. \ref{sec:harmonic}). It has been shown
that the $n^{\rm th}$ order term of the feedback signal effectively
enhances the $n^{\rm th}$ Fourier components of the interaction
function. The time-delay of the $n^{\rm th}$ term is utilized to
arbitrarily tune the balance of even and odd parts of the Fourier
components.  These correspondences appear intuitively reasonable, as the
$n^{\rm th}$ power of the harmonic signal makes a component of harmonic
signal having $n$ times frequency and the time delay shifts the
waveform. In general, the higher order harmonics in the
interaction function are responsible for complex dynamical behavior
including dynamical clustering. Our methodology provides a framework for
tuning all the harmonics in the interaction function.

A major advantage of our methodology is that the feedback resulting in a
target interaction function can be designed through the knowledge of the
macroscopic observables of an isolated oscillator, that is, the waveform
and the phase response function. When focusing on engineering
synchronization properties, a microscopic investigation of the system is
not needed.  This point is beneficial when applications to biological
systems are considered.  It is usually a formidable task to construct an
appropriate, detailed mathematical model of a biological
system. However, the investigation of the phase response function is
often possible; the PRC's of circadian oscillators with respect to light
or temperature stimuli have been extensively measured \cite{johnson99}
as well as the PRC's of neurons with respect to electric stimuli
\cite{galan05,tsubo07}.

Our methodology may be used not only to induce dynamical order but also
to destroy synchronization. In a previous paper \cite{kiss07}, we have
demonstrated that a theoretically designed feedback successfully
desynchronizes a population of chemical oscillators which otherwise
shows simple synchronization due to global interaction among elements.
The model-engineered feedback may find application in pacemaker and
anti-pacemaker design for medical use (tremors, epilepsy).

Because of the robustness of phase description of limit-cycle
oscillators, our methodology for designing interaction functions with
feedback is robust against (at least weak) noise. However, when a
complex dynamical structure is designed, we need to consider the
(structural) stability of the designed dynamical behavior in the
presence of noise. For example, global noise can enhance the extent of
phase synchronization \cite{zhou02}, but can destroy subtle structures
like slow switching \cite{hansel93,kori01}. Therefore, the precision of
the fitted interaction function and the overall gain shall be carefully
chosen in the presence of noise to obtain the desired structure. Because
the proposed methodology was shown to work in the experimental system,
our method should be applicable in systems with weak noise and
well-defined oscillator waveform and response function.

Limitations to our approach should be noted. We have focused on mild
engineering, mild such that essential dynamical properties of elements
are preserved. This strategy allows us to use the phase model. The phase
models cannot be used with strong feedback because of amplitude effects.
Also, the applicability of our method to chaotic
oscillators is unclear because the rigorous phase description for
chaotic oscillators has not been established yet. In coupled chaotic
oscillators, various types of collective behavior arise and some of them
are analogous to those in weakly coupled limit-cycle oscillators, such
as phase synchronization \cite{pikovsky01,boccaletti02}. It would be
thus worth trying to extend the method to chaotic oscillators.  Another
issue arises in cases where limit-cycle oscillators have inherent
complex interactions. In the present paper, oscillators are assumed to
be independent (i.e., uncoupled) unless feedback is applied. In the
presence of inherent global coupling, we have shown the
desynchronization is possible using our methodology \cite{kiss07}.  What
happens if the oscillators are coupled via space dependent interactions
or complex networks? This issue requires further
exploration, for example, in chemical reaction-diffusion systems
\cite{kobayashi08} and control neural networks.

\acknowledgments We thank Alexander Mikhailov, Yoshiki Kuramoto, Ichiro
Tsuda, and Yasumasa Nishiura for useful discussions and warm
hospitality. H. K. acknowledges financial support from the Alexander von
Humboldt Foundation and from the 21st Century COE Program ``Mathematics
of nonlinear structures via singularities'' in Hokkaido University,
Japan. This work was supported in part by the National Science
Foundation under grant CBET-0730597.

\appendix

\section{existence and stability of the balanced cluster states} \label{sec:cluster_state}
The balanced $n$ cluster state may be described as 
\begin{equation}
 \phi_{j \in C_k} = \Omega t + 2 k \pi/n,
  \label{balanced}
\end{equation}
where the set $C_k$ identifies the oscillators forming the cluster $k$
($k=0,\ldots,n-1$) and each set includes $N/n$ elements.  Such a solution
always exists in the phase model (\ref{pm}). Substituting
(\ref{balanced}) into Eq. (\ref{pm}), we obtain
\begin{equation}
 \Omega = \frac{K}{n} \sum_{k=0}^{n-1} H(2 k \pi/n) =
  H_0+2 \sum_{l=1}^{\infty} {\rm Re} H_l.
\end{equation}
The linear stability problem for the balanced cluster states has been
studied by Okuda \cite{okuda93}. The eigenvalues were found to be
\begin{equation}
 \lambda_{\rm intra}^{(n)} = - 2 \sum_{l=1}^{\infty} l {\rm Im} H_{nl},
 \label{Stab_1}
\end{equation}
\begin{equation}
 {\rm Re} \lambda_{{\rm inter},p} = \lambda_{\rm intra}^{(n)} -
  \sum_{l=1}^{\infty} l {\rm Im} \{ H_{n(l-1)+p} + H_{nl-p} \},
  \label{Stab_2}
\end{equation}
\begin{equation}
 \lambda_0= 0
\end{equation}
where $\lambda_{\rm intra}^{(n)}$ is associated with intra-cluster fluctuations
($N-n$ multiplicity), $\lambda_{{\rm inter},p}$ ($p=1,\ldots,n-1$) are
associated with inter-cluster fluctuations, and $\lambda_0$ is
associated with the identical phase shift.

For the interaction function with ${\rm Im} H_n>0$ and ${\rm Im} H_l
\leq 0$ for $l \neq n$, the following relation holds:
y\begin{equation}
 {\rm Re} \lambda_{{\rm inter},p} < \lambda_{\rm intra}^{(n)} \equiv
  \lambda^{(n)}_{\rm max}.
\end{equation}
Thus, the $n$ cluster state is linearly stable if and only if $\lambda_{\rm
max}^{(n)}<0$.


\end{document}